\documentstyle[11pt,a4,epsfig]{article}
\newcommand{\Pslash}{P \hspace{-0.24cm} / \,}
\newcommand{\kslash}{k \hspace{-0.2cm} / \,}
\newcommand{\nslash}{n \hspace{-0.2cm} / \,}

\begin{document}

\title{\bf On the validity of Lorentz invariance relations between
parton distributions\footnote{Talk given at the 10$^{th}$ Workshop on High Energy
Spin Physics, Dubna, Russia, September 2003}}

\author{M.~Schlegel, A.~Metz
 \\[0.3cm]
{\it Institut f\"ur Theor. Physik II, Ruhr-Universit\"at Bochum, 
     D-44780 Bochum, Germany}}

\date{\today}
\maketitle 

\begin{abstract}
\noindent
Lorentz invariance relations connecting twist-3 parton distributions
with transverse momentum dependent twist-2 distributions have been
proposed previously. 
These relations can be extracted from a covariant decomposition of 
the quark-quark correlator. 
It is argued, however, that the derivation of the Lorentz invariance 
relations fails if the path-ordered exponential is taken into account 
in the correlator. 
The model independent analysis is supplemented by an explicit calculation 
of the corresponding parton distributions in perturbative QCD with a 
quark target, and in a simple spectator model.
We also clarify the status of a specific calculation of time-reversal 
even parton distributions in light-cone gauge.
\end{abstract}

\section{Introduction}
Transverse-momentum dependent ($k_{\perp}$-dependent) parton distributions 
and twist-3 parton distributions play an important role in describing various 
hard processes like semi-inclusive deep inelastic scattering or the 
Drell-Yan process. 
In particular, these functions enter certain spin and/or azimuthal 
asymmetries~\cite{jaffe_91,levelt_94}, and recently results for such
asymmetries have been reported by the HERMES and CLAS 
collaborations~\cite{HERMES_00,CLAS_03}.

In Refs.~\cite{tangerman_94,mulders_96,boer_98} so-called Lorentz invariance 
relations (LI-relations) were introduced, which connect twist-3 distributions
with moments of $k_{\perp}$-dependent twist-2 distributions. 
The LI-relations impose important constraints on the distribution functions,
which allow one to eliminate unknown structure functions in favor of
known ones. 
Moreover, these relations could be very useful for deriving the evolution 
of moments of $k_{\perp}$-dependent distribution functions~\cite{henneman_02}.

Here, we argue that the LI-relations are violated because in their 
derivation a dependence on a light-cone vector was neglected~\cite{goeke_03}.
Our model independent analysis is supplemented by explicit model results.
Two specific LI-relations for time-reversal even (T-even) distributions
have already been questioned in~\cite{kundu_01}, where the involved
functions were computed in light-front Hamiltonian QCD using a dressed
quark target.
However, the treatment in Ref.~\cite{kundu_01} is not fully gauge invariant,
which motivated us to revisit the calculation.

\section{Model independent analysis}
The derivation of the LI-relations in Refs.~\cite{tangerman_94,mulders_96,boer_98}
is based on the consideration of the quark-quark correlator for the nucleon
(characterized by its momentum $P$ and a spin vector $S$)
\begin{equation} \label{unintegriert}
\Phi_{ij}(P,k,S|{\rm path}) \equiv 
\frac{1}{(2\pi )^{4}} \int d^{4}\xi \; e^{ik \cdot \xi}
\left\langle P,S | \bar{\psi }_{j}(0) {\mathcal{W}}(0;\xi|{\rm path})
             \psi _{i}(\xi) | P,S\right\rangle ,
\end{equation}
where the Wilson line ensuring gauge invariance is given by
\begin{equation}
{\mathcal{W}}(0;\xi|{\rm path}) = 
{\mathcal{P}} exp\left\{ -ig \int _{0}^{\xi }ds^{\mu }\; A_{\mu }(s)\right\} _{{\rm path}}.
\end{equation}
Note that the correlator in (\ref{unintegriert}) doesn't enter a factorization 
theorem for a physical process.
Taking into consideration the constraints due to hermiticity and parity the 
correlator in Eq.~(\ref{unintegriert}) can be decomposed in the most general
Lorentz invariant way according to
\begin{equation} \label{decomp}
\Phi_{ij}(P,k,S|{\rm path}) = M \, A_{1} + \Pslash \, A_2 
+ {\kslash} \, A_{3} + \frac{i}{2M} [{\Pslash},{\kslash}] \, A_{4} + \ldots ,
\end{equation}
where the $A_{i} = A_{i}(k^{2},k \cdot P)$ are unknown coefficient functions.
We have limited the list of structures to those relevant for an unpolarized
target.
The factors of the nucleon mass $M$ in (\ref{decomp}) were introduced in
order that all $A_i$ have the same dimension.

On the other hand, the ($k_{\perp}$-dependent) parton distributions entering 
the hadronic part of physical processes are defined through a correlation 
function $\Phi_{ij}(x,k_{\perp},S|{\rm path})$ by taking projections with 
appropriate Dirac matrices.
This correlator is connected to the one in (\ref{unintegriert}) by means of
\begin{equation} \label{reduce}
\Phi_{ij}(x,k_{\perp},S|{\rm path}) = 
\int dk^- \Phi_{ij}(P,k,S|{\rm path}) \Big|_{k^+ = x P^+} .
\end{equation}
The path on the {\it lhs} in (\ref{reduce}) is fixed when doing a proper 
factorization, and will be specified below.
The path on the {\it rhs} has to be chosen such that after the $k^-$-integration 
it matches with the one on the {\it lhs}.
Eq.~(\ref{reduce}) immediately relates the parton distributions with the 
amplitudes $A_i$.
One finds, e.g.,
\begin{eqnarray}
f_{1}(x,\vec{k}_{\perp}^2) & = & 
 2 P^{+} \int dk^{-} \; \Big( A_{2}+xA_{3} \Big) ,
\\
h_{1}^{\perp}(x,\vec{k}_{\perp}^2) & = & 
 2 P^{+} \int dk^{-} \; \Big( -A_{4} \Big) ,
\\
h(x,\vec{k}_{\perp}^2) & = & 
 2 P^{+} \int dk^{-} \; \bigg( \frac{k \cdot P - xM^{2}}{M^{2}} A_{4} \bigg) ,
\end{eqnarray}
where $f_1$ is the usual unpolarized quark distribution, while $h_1^{\perp}$ and
$h$ are T-odd distributions~\cite{mulders_96,boer_98}.
Comparing now these expressions for the parton distributions, one
can find LI-relations between those distributions which contain the 
same $A_{i}$'s. 
We list here the most important ones~\cite{tangerman_94,mulders_96,boer_98},
\begin{eqnarray}
g_{T}(x) & = & 
 g_{1}(x) + \frac{d}{dx} g_{1T}^{(1)}(x),
 \label{g-LI} \\
h_{L}(x) & = & 
 h_{1}(x) - \frac{d}{dx} h_{1L}^{\perp (1)}(x),
 \label{hl-LI} \\
f_{T}(x) & = & 
 -\frac{d}{dx} f_{1T}^{\perp (1)}(x),
 \label{f-LI} \\
h(x) & = & 
 -\frac{d}{dx} h_{1}^{\perp (1)}(x),
\label{h-LI}
\end{eqnarray}
with 
\begin{equation}
g_{1T}^{(1)}(x) = \int d^2\vec{k}_{\perp} \, \frac{\vec{k}_{\perp}^2}{2M^2} \,
 g_{1T}(x,\vec{k}_{\perp}^2), \quad {\rm etc.}
\end{equation}
All distributions on the {\it lhs} in Eqs.~(\ref{g-LI})--(\ref{h-LI}) are of 
twist-3, whereas the functions on the {\it rhs} appear unsuppressed in 
observables.
The two LI-relations in (\ref{g-LI},\ref{hl-LI}) connect T-even parton 
distributions, the ones in (\ref{f-LI},\ref{h-LI}) contain T-odd distributions.

In Ref.~\cite{goeke_03} we have argued, that the LI-relations are invalid.
The crucial observation is that the decomposition in Eq.~(\ref{decomp}) is 
incomplete because the presence of the gauge link leads to a dependence on an 
additional light-like vector.
This point becomes obvious when keeping in mind the appropriate gauge link 
structure of the correlator 
\begin{equation} \label{integriert}
\Phi_{ij}(x,k_{\perp},S|{\rm path}) =
\frac{1}{(2\pi )^{3}} \int d\xi^- d^2\vec{\xi}_{\perp} \; 
 e^{i(k^+ \xi^- - \vec{k}_{\perp} \cdot \vec{\xi}_{\perp})}
\left\langle P,S | \bar{\psi }_{j}(0) {\mathcal{W}}(0;\tilde{\xi}|{\rm path})
             \psi _{i}(\tilde{\xi}) | P,S\right\rangle ,
\end{equation}
with $\tilde{\xi }=(0,\xi ^{-},\vec{\xi }_{T})$ and the Wilson 
line~\cite{collins_82,collins_02,ji_02,belitsky_03}
\begin{equation} \label{path}
{\mathcal{W}}(0;\tilde{\xi}|{\rm path}) =
{\mathcal{W}}(0,0,\vec{0}_{\perp} ; 0,\infty,\vec{0}_{\perp}) \times
{\mathcal{W}}(0,\infty,\vec{0}_{\perp} ; 0,\infty,\vec{\xi}_{\perp}) \times
{\mathcal{W}}(0,\infty,\vec{\xi}_{\perp} ; 0,\xi^-,\vec{\xi}_{\perp}) \,.
\vphantom{\frac{1}{1}}
\end{equation}
To parameterize this gauge link a light-like vector $n$ (in addition to the
light-cone direction given by the target momentum) is needed, which has to 
show up also in the unintegrated correlator in Eq. (\ref{unintegriert})
due to the connection in (\ref{reduce}). 
Therefore, more covariant structures in (\ref{decomp}) with new coefficient 
functions $B_{i}$ show up~\cite{goeke_03,bacchetta_04}\footnote{The 
last term in the second line in (\ref{corr_full}) has been overlooked in 
Ref.~\cite{goeke_03} as pointed out recently in~\cite{bacchetta_04}. 
This term, however, does not influence the discussion of the LI-relations.}
\begin{eqnarray} \label{corr_full}
\Phi_{ij}(P,k,S|{\rm path}) & = & 
 M \, A_{1} 
 + \Pslash \, A_2 
 + {\kslash} \, A_{3} 
 + \frac{i}{2M} [{\Pslash},{\kslash}] \, A_{4} + \ldots 
\\
& & + \frac{M^2}{P \cdot n} \nslash \, B_1
 + \frac{i M}{2 P \cdot n} [\Pslash,\nslash] \, B_2
 + \frac{i M}{2 P \cdot n} [\kslash,\nslash] \, B_3
 + \gamma_5 \frac{\epsilon^{\mu\nu\rho\sigma}}{P \cdot n} \, 
   \gamma_{\mu} P_{\nu} n_{\rho} k_{\sigma} \, B_4 + \ldots ,
\nonumber
\end{eqnarray}
where again we only listed the structures that appear in the case of an 
unpolarized target.
The new terms typically modify the expressions of the parton distributions 
in terms of the coefficient functions. 
As an example we consider the two functions entering the LI-relation 
in (\ref{h-LI}) for which we now find
\begin{eqnarray}
h_{1}^{\perp}(x,\vec{k}_{\perp}^2) & = & 
 2 P^{+} \int dk^{-} \; \Big( -A_{4} \Big) ,
\\
h(x,\vec{k}_{\perp}^2) & = & 
 2 P^{+} \int dk^{-} \; \bigg( \frac{k \cdot P - xM^{2}}{M^{2}} A_{4} 
                               + \Big( B_2 + x B_3 \Big) \bigg) .
\end{eqnarray}
Obviously, the presence of the amplitudes $B_{2}$ and $B_{3}$ in the expression
for $h$ spoils this particular LI-relation. 
By considering structures which depend on the target polarization the violation 
of the relations in (\ref{g-LI})-(\ref{f-LI}) can be shown as well.

\section{Model calculations}
We now want to supplement our model independent analysis by an explcit
model calculation of the relevant parton distributions.

For the two relations in (\ref{g-LI},\ref{hl-LI}) all involved distributions
have already been computed for a dressed quark target in light-front 
Hamiltonian QCD in Ref.~\cite{kundu_01}, and a violation of the LI-relations 
has been observed.
However, the transverse Wilson line at the light-cone infinity in Eq.~(\ref{path}), 
whose importance for light-cone gauge calculations has been realized only 
afterwards~\cite{ji_02,belitsky_03}, was not taken into account in~\cite{kundu_01}.
As a consequence, the results for the moments of the $k_{\perp}$-dependent 
distributions $g_{1T}$ and $h_{1L}^{\perp}$ in Eqs.~(\ref{g-LI},\ref{hl-LI})
might change in a fully gauge invariant treatment.
\\
We have repeated this pQCD calculation for a quark target in Feynman gauge,
where the Wilson line at the light-cone infinity doesn't contribute.
Only the two parts of the link in Eq.~(\ref{path}) which run along the 
$\xi^-$-direction are relevant.
By introducing intermediate states in the correlator in (\ref{integriert}) one 
can make an expansion up to first order in $\alpha _{s}$,
\begin{eqnarray}
\lefteqn{\Phi_{ij}(x,k_{\perp},S|{\rm path}) = 
\frac{1}{(2\pi )^{3}} \int d\xi^- d^2\vec{\xi}_{\perp} \; 
 e^{i(k^+ \xi^- - \vec{k}_{\perp} \cdot \vec{\xi}_{\perp})}}
\\
& & \mbox{} \times \bigg[ \langle q;P,S,a | \bar{\psi }_{j}(0) 
    {\mathcal{W}}(0,0,\vec{0}_{\perp};0,\infty,\vec{0}_{\perp}) | 0 \rangle \langle 0 | 
    {\mathcal{W}}(0,\infty,\vec{\xi}_{\perp};0,\xi^-,\vec{\xi}_{\perp})
    \psi _{i}(\tilde{\xi}) | q;P,S,a \rangle 
\nonumber \\
& & \qquad + \sum_{r,b} \int \frac{d^3 l}{2(2\pi)^3 E_l}
    \langle q;P,S,a | \bar{\psi }_{j}(0) 
    {\mathcal{W}}(0,0,\vec{0}_{\perp};0,\infty,\vec{0}_{\perp}) | g;l,r,b \rangle 
\nonumber \\
& & \hspace{3.1cm} \mbox{} \times 
    \langle g;l,r,b | {\mathcal{W}}(0,\infty,\vec{\xi}_{\perp};0,\xi^-,\vec{\xi}_{\perp})
    \psi _{i}(\tilde{\xi}) | q;P,S,a \rangle + \ldots \bigg] .
\nonumber 
\end{eqnarray}
Up to ${\mathcal{O}}(\alpha _{s})$ we only need to consider gluons as 
particles in the intermediate state.
The one-loop calculation is UV-divergent, and we regularize the expressions
by calculating in $4 - 2\varepsilon$ dimensions.
The resulting parton distributions are extracted from 
$\Phi _{ij}(x,k_{\perp},S|{\rm path})$ by appropriate projections and integration
over $\vec{k}_{\perp}$. 
For the parton distributions in Eq.~(\ref{g-LI}) we obtain
\begin{eqnarray} \label{g1}
g_{1}(x) & = & \delta (1-x) 
 + \frac{\alpha _{s}}{2\pi } \, C_{F} \, \frac{1+x^{2}}{1-x} \, 
 \frac{1}{\varepsilon} + \ldots \,,
\\ \label{g2}
g_{T}(x) & = & \delta (1-x)
 + \frac{\alpha _{s}}{2\pi } \, C_{F} \, \frac{1+2x-x^{2}}{1-x} \, 
 \frac{1}{\varepsilon} + \ldots \,,
\\ \label{g3}
g_{1T}^{(1)}(x) & = & 
 - \frac{\alpha _{s}}{2\pi } \, C_{F} \, x(1-x) \,
 \frac{1}{\varepsilon} + \ldots \,.
\end{eqnarray}
The dots in Eqs.~(\ref{g1})--(\ref{g3}) indicate virtual radiative corrections
and/or higher orders in $\alpha_s$.
The virtual contributions contain an explicit factor $\delta (1-x)$ 
and remove the singularity for $g_1$ and $g_T$ at $x = 1$ 
(see, e.g., Ref.~\cite{kundu_01}).
In order to check the LI-relations it is sufficient to consider the UV-divergent
part of the real radiative corrections which are proportional to $1/\varepsilon$.
With the results in~(\ref{g1})--(\ref{g3}) one finds that the LI-relation~(\ref{g-LI})
is fulfilled at leading order, but the $\alpha_s$-corrections spoil the relation.
By means of an analogous calculation it can be shown that relation (\ref{hl-LI}) 
is violated as well.
\\
The expressions in~(\ref{g1})--(\ref{g3}) completely agree with the ones 
obtained in Ref.~\cite{kundu_01}, where a UV-cutoff has been used to regulate
the divergent $\vec{k}_{\perp}$-integral.
In particular, the results for $g_{1T}$ and $h_{1L}^{\perp}$ coincide. 
From this observation we conclude that for our specific calculation the 
transverse Wilson line at the light-cone infinity is irrelevant in 
light-cone gauge.
Whether this result holds in general for T-even parton distributions remains 
to be seen.

The situation for the LI-relations (\ref{f-LI},\ref{h-LI}) which involve T-odd 
functions is simpler.
In this case we can use results which already exist in the literature in 
addition to arguments based on T-invariance.
In the framework of a simple diquark spectator model for the nucleon it has 
been shown that $f_{1T}^{\perp}(x,\vec{k}_{\perp})$ doesn't 
vanish~\cite{brodsky_02,collins_02,ji_02}.
On the other hand, the $\vec{k}_{\perp}$-independent parton distribution 
$f_{T}(x)$ is zero because of T-invariance~\cite{collins_93,boer_03a}.
Therefore, the relation (\ref{f-LI}) is violated in this model.
The same reasoning can be used for the LI-relation (\ref{h-LI}):
an explicitly non-vanishing result for $h_{1}^{\perp }(x,\vec{k}_{\perp})$ has 
been obtaind in the diquark spectator model~\cite{goldstein_02,boer_03b},
while T-invariance rules out a non-zero $h(x)$. 
\\
Finally, we mention that by considering the T-odd case we can easily give even 
stronger support to the picture that the presence of the light-cone vector $n$, 
and consequently the presence of the amplitudes $B_i$, is at the origin of the 
violation of the LI-relations. 
If one would define the $k_{\perp}$-dependent parton distributions with a single
straight Wilson line connecting the two quark fields, then neither the correlator
in Eq.~(\ref{integriert}) nor the one in (\ref{unintegriert}) would contain a 
$n$-dependence.
Hence, the second line in (\ref{corr_full}) containing the amplitudes $B_i$ would 
be absent, and the LI-relations should be valid.
For the relations (\ref{f-LI},\ref{h-LI}) one observes readily that this 
expectation is indeed correct, because both sides of the relations would vanish
because of T-invariance.
\\[0.6cm]
\noindent
{\bf Note added:} It is interesting, that the very last $n$-dependent term in 
Eq.~(\ref{corr_full}) gives rise to a new twist-3 T-odd parton 
distribution~\cite{bacchetta_04} and, therefore, to an observable effect.
The spectator model calculations of the beam single spin asymmetry for 
semi-inclusive deep inelastic scattering in Refs.~\cite{afanasev_03,metz_04} 
suggested the existence of this new function. 
\\[0.6cm]
\noindent
{\bf Acknowledgements:}
We thank K.~Goeke, P.V.~Pobylitsa and M.V.~Polyakov for various discussions 
and for the fruitful collaboration which lead to the model independent analysis 
presented here.
The work has been partly supported by the Sofia Kovalevskaya Programme of the
Alexander von Humboldt Foundation, the DFG and the COSY-Juelich project.

\end{document}